\documentclass[aps,prl,showpacs,twocolumn]{revtex4}
\def\ket#1{|#1\rangle}
\def\bra#1{\langle#1|}

\begin{document}
\title{Three-party pure quantum states are determined by two two-party 
       reduced states}
\author{Lajos Di\'osi}
\email{diosi@rmki.kfki.hu}
\homepage{www.rmki.kfki.hu/~diosi} 
\affiliation{
Research Institute for Particle and Nuclear Physics\\
H-1525 Budapest 114, POB 49, Hungary}
\date{\today}

\begin{abstract}
We can uniquely calculate almost all entangled state vectors of 
tripartite systems $ABC$ if we know the reduced states of any two 
bipartite subsystems, e.g., of $AB$ and of $BC$. We construct the 
explicit solution. 
\end{abstract}


\maketitle
Generic multiparty composite quantum states establish complex multiparty
correlations. In the particular case of pure composite states, however,
recent evidences have shown that higher order correlations follow from
lower order ones \cite{LinPopWoo,LinWoo,HanZhaGuo}. Such quantum features 
\cite{qf} are of central interest in the modern field of quantum 
information \cite{NieChu} as well as in the more traditional field of 
many-body physics \cite{Col}. 

Generic three-party pure quantum states have shown to be uniquely 
determined by their two-party reduced states \cite{LinPopWoo,LinWoo}. 
Consider, e.g., a composite pure state 
$\rho_{ABC}=\ket{\psi_{ABC}}\bra{\psi_{ABC}}$ 
of three parties $A,B,C$ of dimensions $d_A,d_B,d_C,$ respectively. 
Let $\rho_{AB},\rho_{BC},\rho_{AC}$ denote the two-party reduced 
states. In case of three qubits, these three reduced states will 
uniquely determine the composite state $\rho_{ABC}$ in almost all 
cases \cite{LinPopWoo}. For higher dimensions, satisfying the 
`triangle-inequality' $d_A\geq d_B+d_C-1$, an alternative theorem 
holds: the two reduced states $\rho_{AB}$ and $\rho_{AC}$ are already 
sufficient to calculate the state $\rho_{ABC}$ of the whole system 
\cite{LinWoo}. Note that in both cases one calculates $\rho_{ABC}$ 
{\it without} assuming that $\rho_{ABC}$ is pure. It comes out from the 
reduced states. If one assumes it then a stronger statement holds. 
As we shall prove in the present work, almost all pure composite states 
$\ket{\psi_{ABC}}$ can uniquely be calculated from the knowledge of 
any two of the two-party reduced states if one knows already that 
$\rho_{ABC}$ is pure. This result holds in any finite dimensions. 
We present explicit equations for $\ket{\psi_{ABC}}$.
  
For concreteness, let us prove how a generic $|\psi_{ABC}\rangle$
is determined by $\rho_{AB}$ and $\rho_{BC}$. Obviously, the latters
determine the three single-party reduced states $\rho_A,\rho_B,\rho_C$
as well. One shall diagonalize them, e.g.:
\begin{equation}
\rho_A=\sum_i p_A^i\ket{i}\bra{i},~~~p_A^i>0.
\end{equation}
Similarly, $\ket{j}$ and $\ket{k}$ stand for the eigenvectors with 
non-zero eigenvalues $p_B^j,p_C^k$ of $\rho_B$ and $\rho_C$ 
respectively. Since $\rho_{ABC}$ is pure, the reduced state $\rho_A$ 
shares its eigenvalues $p_A^i$ with $\rho_{BC}$: 
\begin{equation}
\rho_{BC}=\sum_i p_A^i\ket{i;BC}\bra{i;BC},
\end{equation}
where $\ket{i;BC}$ are the orthogonal eigenvectors of $\rho_{BC}$ with
non-zero eigenvalues. Similarly, we introduce the orthogonal 
decomposition of $\rho_{AB}$ as well, with non-zero eigenvalues $p_C^k$ 
and eigenvectors $\ket{k;AB}$. We may omit decomposition of $\rho_{AC}$: 
it is not required by the present proof. 
From the spectral decompositions (1,2)
we can reconstruct the Schmidt-decomposition of {\it all} three-party 
pure states compatible with $\rho_A$ and $\rho_{BC}$:
\begin{equation}
\ket{\psi_{ABC};\alpha}
=\sum_i \exp({\rm i}\alpha_i)\sqrt{p_A^i}\ket{i}\otimes\ket{i;BC},
\end{equation}
where $\alpha\equiv\{\alpha_i\}$ is the set of phases to be specified 
later. From the spectral decompositions of $\rho_{AB}$ and $\rho_C$, 
we have another family of {\it all} pure states compatible with 
$\rho_{AB}$ and $\rho_C$:
\begin{equation}
\ket{\psi_{ABC};\gamma}
=\sum_k \exp({\rm i}\gamma_k)\sqrt{p_C^k}\ket{k;AB}\otimes\ket{k}.
\end{equation}
Since the true $\ket{\psi_{ABC}}$ is compatible with both $\rho_{AB}$ 
and $\rho_{BC}$ (and thus with $\rho_A,\rho_C$) therefore at least one 
solution {\it exists} for the $\alpha_i$'s and $\gamma_k$'s such that
\begin{equation}
\ket{\psi_{ABC};\alpha}=\ket{\psi_{ABC};\gamma}.
\end{equation}
We are going to prove that this solution is unique hence the state (5),
derived from $\rho_{AB}$ and $\rho_{BC}$, will be the true 
$\ket{\psi_{ABC}}$. 

First we cast the vectorial equation (5) into equations for amplitudes.
Let us calculate the following coefficients:
\begin{equation}
{\cal A}^i_{jk}=\langle jk\ket{i;BC},~~~
{\cal C}^k_{ij}=\langle ij\ket{k;AB}.
\end{equation}
They are non-vanishing for a generic state $\ket{\psi_{ABC}}$. 
In fact, the eigenvectors (with non-zero eigenvalues) of
a composite state are superpositions of the direct-products formed by 
the eigenvectors (with non-zero eigenvalues) of the respective
subsystem reduced states \cite{sup}. In our case, we use the following 
expansions:
\begin{equation}
\ket{i;BC}=\sum_{jk}{\cal A}^i_{jk}\ket{jk},~~~
\ket{k;AB}=\sum_{ij}{\cal C}^k_{ij}\ket{ij}.
\end{equation}
Substituting them into eqs.(3) and (4), considering
orthogonality of the product states $\ket{ijk}$, we expand eq.(5) into 
the following set of compatibility equations between $\alpha$
and $\gamma$: 
\begin{equation}
\exp({\rm i}\alpha_i)\sqrt{p_A^i}{\cal A}^i_{jk}=
\exp({\rm i}\gamma_k)\sqrt{p_C^k}{\cal C}^k_{ij}
\end{equation}
for all $i,j,k$. Multiplying the l.h.s by the complex conjugate 
of the r.h.s. and the r.h.s. by the complex conjugate of the l.h.s.
will cancel the factors $\sqrt{p_A^i p_C^k}$, yielding:
\begin{equation}
\exp[{\rm i}(\alpha_i-\gamma_k)]
{\cal A}^i_{jk} \overline{{\cal C}}^k_{ij}=
\exp[-{\rm i}(\alpha_i-\gamma_k)]
\overline{{\cal A}}^i_{jk} {\cal C}^k_{ij}.
\end{equation}
Finally we obtain the following simple equations:
\begin{equation}
\alpha_i-\gamma_k={\rm arg}\sum_j \overline{{\cal A}}^i_{jk}{\cal C}^k_{ij}
\end{equation}
for all $i$ and $k$. The solution ${\alpha_i,\gamma_k}$ is then trivial and 
unique upto an (irrelevant) constant phase shift 
$\alpha_i\rightarrow\alpha_i+\chi,\gamma_k\rightarrow\gamma_k+\chi$. 
The constant $\chi$ contributes to an irrelevant phase 
factor $\exp({\rm i}\chi)$ in front of the pure state (5).

Ref.\cite{LinWoo} considers the generic pure state $\ket{\psi}$ of a large 
number of identical parties of dimension $d$ each. The authors 
derived the upper bound $\alpha_U=2/3$ on the fraction of parties 
whose reduced states enable one to reconstruct $\ket{\psi}$. 
The lower bound $\alpha_L=1/2$ was obtained for large $d$. 
It should be observed that, for the lower bound, the authors assume 
that one restricts the reconstruction for pure states. For such
conditions, our alternative theorem will sharpen the upper bound 
$\alpha_U=2/3$. Let us group the parties into three subsystems $A,B,C$ 
where, e.g., $A$ is a single $d$-state system while $B$ and $C$ share on 
the rest equally or almost equally. According to our theorem, $\rho_{AB}$ 
and $\rho_{AC}$ determine a generic pure state $\ket{\psi}$ of the whole 
system. This yields $\alpha_U=1/2$ asymptotically. Observe the coincidence 
with the lower bound $\alpha_L=1/2$ \cite{LinWoo}. Accordingly, there must
be an (almost) one-to-one mapping between the space of pure states of the 
whole system  and (a certain region in) the space of the reduced states of 
all fractions $\sim1/2$ of the whole.

We summarize the steps reconstructing a state vector $\ket{\psi_{ABC}}$ 
from two density matrices $\rho_{AB}$ and $\rho_{BC}$. First we calculate
$\rho_A,\rho_B,\rho_C$. Then we diagonalize 
$\rho_{AB},\rho_{BC},\rho_A,\rho_B,\rho_C$, and calculate the
coefficients ${\cal A}^i_{jk},{\cal C}^k_{ij}$ (6). The wanted pure state 
$\ket{\psi_{ABC}}$ takes the form (3) with
\begin{equation}
\alpha_i={\rm arg}\sum_j \overline{{\cal A}}^i_{jk}{\cal C}^k_{ij}~~,
\end{equation}
where $k$ is set to any fixed value. Recall that the
${\cal A}^i_{jk}$'s and ${\cal C}^k_{ij}$'s are not independent at all.
The above particular expression of $\alpha_i$ could well be replaced
by a variety of equivalent expressions of them. One can, e.g., take
any fixed value for $j$ instead of the summation over $j$ which we
took for representation invariance. To display explicit invariance of
the full solution we need further investigations on the underlying 
geometric structure.

Finally we mention a possible extension of the method for spatial 
tomography. Assume that we have to reconstruct a spatial wave function
$\psi(xyz)$ from planar projections. Let us define the density
matrices in the $XY$- and $YZ$-planes, e.g.:
\begin{equation}
\rho_{XY}(xy;x'y')=\int\psi(xyz)\overline{\psi}(x'y'z)dz,
\end{equation}
and a similar equation for $\rho_{YZ}$.
If our theorem remains valid for infinite dimensions as well then
the reduced states $\rho_{XY},\rho_{YZ}$ would determine the original 
spatial wave function $\psi(xyz)$. The choice of concrete equations of 
reconstruction would then require special care against numeric 
instabilities.


\begin{thebibliography}{99}
\bibitem{qf} In classical multiparty systems the corresponding statement
would be trivial since pure states are zero-uncertainty states
described by multivariate delta-functions.
\bibitem{sup} One proves that 
${\cal S}(\rho_{AB})\subseteq {\cal S}(\rho_A)\otimes {\cal S}(\rho_B)$
where ${\cal S}(\rho)$ is the {\it support} of $\rho$. The relationship
is trivial for pure $\rho_{AB}$ like, e.g., 
$\rho_{AB}^k=\ket{k;AB}\bra{k;AB}$. Then, for mixed $\rho_{AB}$, 
we can write: ${\cal S}(\rho_{AB})=\bigoplus_k {\cal S}(\rho_{AB}^k)
\subseteq\bigoplus_k [{\cal S}(\rho_A^k)\otimes{\cal S}(\rho_B^k)]
\subseteq\bigoplus_{k,k'}[{\cal S}(\rho_A^k)\otimes{\cal S}(\rho_B^{k'})]
={\cal S}(\rho_A)\otimes{\cal S}(\rho_B)$.
\end{thebibliography}
\end{document}